\documentclass{JHEP}
\usepackage{epsfig}
\preprint{ {\tt hep-th/0105184} }
\newcommand{\be}{\begin{equation}}
\newcommand{\ee}{\end{equation}}
\newcommand{\bea}{\begin{eqnarray}}
\newcommand{\eea}{\end{eqnarray}}
\newcommand{\eq}[1]{(\ref{#1})}

\title{Excitations on wedge states and on the sliver}
\author{Justin R. David \\ Department of Physics, University of
California,\\Santa Barbara, CA 93106, USA.\\
\email{justin@vulcan.physics.ucsb.edu} }
\abstract{ 
We study ghost number one
excitations on the sliver to investigate the solution of string
field actions around the tachyon vacuum. The generalized gluing and
resmoothing theorem is used to develop a method for evaluating the
effective action for excitations on both the wedge states and the
sliver state. 
We analyze the discrete symmetries of the resulting effective action
for excitations on the sliver.
The 
gauge unfixed effective action till level two excitations on the
sliver is evaluated.
This is done for the case with
the BRST operator $c_0$ and $c_0 + (c_2 + c_{-2})/2$ with
excitations purely in the ghost sector. 
We find that the values of the effective potential at the local
maximum
lie close by for the zeroth and the second level 
of approximation.
This indicates that level
truncation in string field theory around the tachyon vacuum using
excitations 
on the sliver converges for both choices of the
BRST operator. It also provides evidence for the conjectured string
field theory actions around the tachyon vacuum.
}

\begin{document}

\section{Introduction}

The phenomenon of tachyon condensation has proved to be a valuable
guide in exploring various open string field theories. 
In this paper we will restrict our focus to the cubic string field
theory of Witten \cite{wit}.
The tachyon in
open bosonic string theory corresponds to the instability of the
D25-brane to decay to the vacuum. 
It has been conjectured by Sen \cite{senc1, senc2}. 
that there exists a  stationary point in
the tachyon potential
\footnote{For earlier studies of the tachyon in open
string theory see \cite{bark}.}. 
The value of the tachyon potential at this
stationary point should agree with the tension of the D25-brane. 
A co-dimension $p$-lump solution is conjectured to represent a
D$(25-p)$-brane in the same theory.
These
conjectures have been verified to a remarkable degree of accuracy
in cubic string field theory using level truncation as an
approximation scheme [5-26]. The D25-brane decays to the vacuum 
when the tachyon
condenses. Thus there should be no physical 
open string excitation around
the tachyon vacuum. Recently strong numerical 
evidence for this conjecture was
found in cubic string field theory in \cite{eltay}. For a review of
the study of tachyon condensation in bosonic and superstring field
theories and other references see \cite{ohmo}.

Lack of an analytical solution for the stable vacuum in cubic string
field theory has made it difficult to study excitations around the
vacuum. In \cite{rsz1} a class of string field theories were put
forward as candidates for string field theory around the tachyon
vacuum. These theories differ from the conventional cubic string
theory of Witten in their choice of the BRST operator. In Witten's
cubic string field theory, the BRST operator $Q_B$ is the usual one
made of the ghost and matter stress energy tensor along with the 
world sheet $c$
and $b$ ghosts. In the candidates 
put forward in \cite{rsz1} the BRST
operator $Q$ was chosen to be solely made of ghosts. Similar actions
were derived from the purely cubic string field theory previously in
\cite{hjmw}. 
The BRST operators considered in \cite{rsz1} has the
advantage of having no manifest physical states. The BRST operators
chosen had trivial cohomology. These operators also preserved the gauge
invariance of the cubic action. 

In \cite{rsz2} classical solutions of  string field theory  around the
tachyon vacuum was studied. These solutions factorised as a tensor
product of  the ghost
sector and the matter. Assuming factorization the matter
sector of the string field equation reduces to
\be
\label{msect}
\Psi* \Psi = \Psi.
\ee
Here $\Psi$ is a string field and the star product is taken over the
matter sector. Recently generalizations to the solution of the above
equation was constructed using the projection operator method in
\cite{rsz3, grotay, kawoku}
\footnote{It is interesting to note that this equation
and its solution using the projection operator method 
has appeared earlier in the context of $c=1$ string field theory in 
\cite{dmw}.}. In \cite{rsz4} solutions to \eq{msect} was constructed
for a general boundary conformal field theory. 
The translational invariant solution to \eq{msect} was constructed using
the explicit representations of the star product in terms of Neumann
functions \cite{gj,sam, ohta} in \cite{kp}.
The matter sector solutions reproduced the ratio of
D-brane tensions to a high degree of accuracy. 

In this paper we explore the the solution of the string field
equations around the tachyon vacuum including the ghost sector. 
We focus on the translational invariant solution which should correspond
to the D25-brane. 
The translational
invariant solution of \eq{msect} was shown to be the matter
sector of the sliver state constructed in \cite{rz}
This prompts us to  study ghost number one excitations on 
the sliver state with the hope of solving the string field equation
around the tachyon vacuum. 
Assuming the solution of the string field theory equation
corresponding to the D25-brane factorizes to the matter and the ghost
sector 
the excitations can be chosen to be purely in the  ghost sector. 
To
evaluate the effective action of these excitations we need to evaluate
the Witten vertex and the quadratic term of these excitations. 
The sliver state is given by
\bea
\label{sliver}
|\Xi\rangle &=& 
\exp (-\frac{1}{3}L_{-2} + \frac{1}{30}L_{-4}
-\frac{11}{1890}L_{-6} + \frac{34}{467775}L_{-8} +\cdots)|0\rangle,
\\ \nonumber
&=& U|0\rangle.
\eea
Here $L_n$'s are the combined matter and ghost Virasoro generators.
The sliver state consists of infinite number of levels. 
Conventional methods
of evaluating the action is therefore difficult. 

We use the generalized gluing
and resmoothing theorem \cite{lpp, sensh}
to develop a method for evaluating the effective
action of excitations on the sliver state \footnote{The author
thanks Ashoke Sen for emphasizing the use of this theorem and pointing
out the reference \cite{sensh}}. 
We mainly analyse excitations of the form 
\be
\tilde{\Phi}(0) = U  \Phi(0) U^{-1} 
\ee
on the
sliver state. Here $U$ is defined in \eq{sliver} and  $\Phi(0)$ is the
string field corresponding to the state $|\Phi\rangle$.
The sliver state is a limit of the wedge state $|n\rangle$ as $n
\rightarrow \infty$. 
Using the generalized gluing and resmoothing theorem 
we first develop the method 
to evaluate the Witten vertex
and the kinetic term for excitations on the
wedge state $|n\rangle$.  Then we take
the limit $n\rightarrow\infty$ to obtain the the Witten vertex and the
kinetic term for excitations above the sliver state. We indicate when
the limit is well defined.

We study the effective action of the 
excitations $\tilde{\Phi}(0)$ on the sliver state
using both $Q= c_0$
and $Q= c_0 + (c_2 +c_{-2})/2$. 
These choices of BRST operators represent equivalent classes of
BRST operators by field redefinitions of the kind $Q\sim e^K Q
e^{-K}$. Here $K$ are conformal transformations which leave the string
midpoint fixed. The operator $c_0$ denotes an equivalent class which
does not annihilate the identity of the star 
algebra, while the operator 
$c_0 + (c_2+ c_{-2})/2$ does.
We verify that the discrete 
symmetry properties of the effective action for
excitations on the sliver state is inherited from the 
cubic action.
We then evaluate the gauge unfixed action for fields till  level $2$ in
$\Phi(0)$  and analyze its local maxima. 
The values of the effective potential at the local maximum lie close
together for both the zeroth and second level approximation for
the two choices of the BRST operator. 
This indicates that level truncation using excitations on
the sliver state converges for both choices of the BRST operator. It
is of interest to point out that level truncation in the string field
theory around the tachyon vacuum with $Q= c_0$ seemed to push the
maximum to zero \cite{rsz1}.

This paper is organized as follows. In section 2 we review the
definition of the wedge states and the sliver state. In section 3 we
review the generalized gluing and resmoothing theorem and use it
obtain the product law of wedge state $|n\rangle *|m\rangle =
|n+m-1\rangle$. In section 4 we use the 
generalized gluing and resmoothing theorem
to obtain the kinetic term and the Witten vertex for excitations
on the wedge states. Section 5 contains the quadratic term and the
Witten vertex for excitations on the sliver state. 
The discrete symmetries of the effective action for excitations are
analyzed.
Then we analyze the effective action for both $Q=c_0$  and
$Q=c_0 + (c_2 + c_{-2})/2$ till level $2$ in $\Phi$.
In section 6 we indicate the difficulty in generalizing  
the Witten vertex and the kinetic
term for other kind of excitations. Section 7 contains our
conclusions. 
Appendix A contains details of the expansion coefficients needed to
write the sliver state in terms of ghost and matter oscillators. 
Appendix B contains the conformal
transformation of non-primary fields involved in the 
effective action.

\section{The wedge states and the sliver state}

In this section we will review the definition of the wedge states
introduced in \cite{rz}. The sliver state is then obtained as a limit
of the wedge state. We construct the sliver sate in terms of 
matter and ghost oscillators as a squeezed state. 

\subsection{The definition of wedge states}

In \cite{rz} the wedge states were defined as conformal transformation
on the $SL(2, R)$ vacuum.
\be
\langle n| = \langle 0 | U_{W_n},
\ee
Where  $\langle 0|$ stands for the left $SL(2, R)$ vacuum and 
$U_{W_n}$ denotes the conformal transformation corresponding to the map
\be
W_n (z) = M\left[ \left( \frac{1+iz}{1-iz} \right)^{\frac{2}{n}}
\right].
\ee
$M$ stands for any $SL(2, C)$ transformation which maps the unit
circle to the real line. For instance we can take
\be
\label{sl2r}
M(z) = -i \frac{z-1}{z+1}.
\ee
The map $[(1+iz)/(1-iz)]^{2/n}$ takes the region inside the 
upper half disc
$|z|\leq 1$ to a wedge of angle $2\pi/n$. Hence the name wedge state.
Given the above $SL(2, C)$ map $M(z)$ the function $W_n(z)$ is given
by $\tan(\frac{2}{n}\tan^{-1}(z))$. For the rest of the paper we will
assume $M(z)$ is given by \eq{sl2r}.

The operator $U_{W_n}$ corresponding to the map $W_n$ is written as
\be
U_{W_n} = \exp(v_0L_0) \exp ( \sum_{n\geq 1} v_n L_n).
\ee
Here $L_n$ are the combined matter and ghost Virasoro generators. 
The coefficients $v_n$ are obtained by comparing coefficients of
different powers of $z$ on both sides of the equation
\be
f(z) = 
\exp\left(\sum_{n\geq 1} v_n z^{n+1}\partial_z\right) 
\exp(v_0z\partial_z ) z.
\ee
Note that this function leaves the origin fixed. The ket $|n\rangle$
is given by
\be
|n\rangle = U^{\dagger}_{W_n} |0\rangle,
\ee
where $U^{\dagger}_{W_n}$ is the BPZ conjugate of $U_{W_n}$. 
It is easy to show that $U^{\dagger}_{W_n}= U_{I\circ W_n^{-1}\circ I}$.
Here $I$ denotes the $SL(2, R)$ transformation $I(z) = -1/z$.
The map $I\circ W_n^{-1} \circ I$ leaves the point at infinity fixed.
Therefore we can write \footnote{We use the symbol $\circ$ to denote
composition of maps as well as the action of a conformal
transformation on a field.}
\be
I\circ W_n^{-1} \circ I(z) = 
\exp\left(\sum_{n\leq -1} v'_n z^{n+1}\partial_z\right) 
\exp(v'_0z\partial_z ) z.
\ee
Once the coefficients $v'_n$ are determined from the above expansion
the operator $U_{I\circ W_n^{-1}\circ I}$ can be constructed as
\be
U_{I\circ W_n^{-1}\circ I} = 
\exp ( \sum_{n\leq -1} v_n L_n)
\exp(v_0L_0).
\ee

The wedge state $|1\rangle$  can be identified with the identity of
the star algebra, while the state $|2\rangle$ is the $SL(2, R)$
vacuum.
It was shown in \cite{rz} that the wedge states obey the product law
\be
\label{pr}
|n\rangle *|m\rangle = |m+n-1\rangle.
\ee
We will re-derive this law by using the 
generalized gluing and resmoothing theorem
explicitly. The sliver state is given as the limit
\be
\label{slylim}
|\Xi\rangle = \lim_{n\rightarrow \infty} |n\rangle.
\ee
In terms of the explicit representation of the operator 
$U_{I\circ W_\infty^{-1}\circ I}$, 
there is a smooth limit given by 
\be
\label{sly}
|\Xi\rangle= 
U_{I\circ W_{\infty}^{-1}\circ I} |0\rangle =   
\exp (-\frac{1}{3}L_{-2} + \frac{1}{30}L_{-4}
-\frac{11}{1890}L_{-6} + \frac{34}{467775}L_{-8} +\cdots)|0\rangle.
\ee
Note that the dependence on $n$ drops out \cite{rz}.
As the operator $U_{I\circ W_{\infty}^{-1}\circ I}$ 
operator commutes with the momentum 
the sliver state is a translational invariant state.
Thus formally from \eq{pr} and  \eq{slylim} the sliver state satisfies
\be
\label{slp}
|\Xi\rangle *|\Xi\rangle = |\Xi\rangle.
\ee

\subsection{Construction of the sliver state in terms of
oscillators}

It is useful to obtain the representation of the sliver state in terms
of the matter and ghost oscillators. From \eq{sly} it is easy to see
that the sliver state is created from the $SL(2, R)$ vacuum
by exponentiation of an operator which is quadratic in the 
oscillators.
Thus it is similar to a squeezed state.
Let the translational invariant 
sliver state be expressed as a squeezed state by
\be
\label{sq}
\langle \Xi | = \langle 0 | \exp \left(
-\frac{1}{2}\eta_{\mu\nu}
\alpha_{m}^{\mu} S_{mn} \alpha_{n}^\nu
+ c_{s} \tilde{S}_{si}b_{i} \right) {\cal N},
\ee
where $m, n = 1\ldots,  \infty$, $i = 2, \ldots, \infty$ , $s = -1
\ldots, \infty$ and $\mu, \nu = 0, \ldots 25$. ${\cal N}$ is a
normalization constant which can be fixed from the equation
\eq{slp}.
We will review the formulae to obtain the width 
matrix  $S$ of the squeezed state,
and then extend that method to obtain the ghost sector width
$\tilde{S}$.
From \eq{sq} we see that 
\bea
\label{cor1}
S_{mn} &=& -\frac{1}{mn}\langle \Xi| \alpha_{-m}^{1}\alpha_{-n}^{1} 
c_{-1} c_0 c_1 |0\rangle, \\ \nonumber
&=& -\frac{1}{mn}\langle 0|U_{W_\infty} 
\alpha_{-m}^1\alpha_{-n}^1 c_{-1}c_0c_1  U^{-1}_{\infty}|0\rangle.
\eea
Here we have used $U^{-1}_{W_\infty}|0\rangle =|0\rangle$ 
the definition of the sliver state and the following commutation
relation for the matter oscillators.
\be
[\alpha_n^\nu, \alpha_m^\mu] = -n \eta^{\nu\mu}\delta(n+m).
\ee
Similarly the width of the ghost sector is given by
\bea
\label{cor2}
\tilde{S}_{si}&=& \langle\Xi | c_{-i}b_{-s} c_{-1}c_0c_1 |0\rangle, 
\;\;\;\;\;\;\;\;\;\;\;  \hbox{     for     } i\geq2, s\geq 2
\nonumber \\
&=& \langle 0 |U_\infty 
c_{-i}b_{-s} c_{-1}c_0c_1 U_\infty^{-1} |0\rangle.
\eea
Here we have used the anti-commutation relation $\{ c_n, b_m\} = \delta
(n+m)$. For $s=-1, 0, 1$ and $i\geq 2$ the width $\tilde{S}$ is given
by
\be
\label{cor3}
\tilde{S}_{-1i} = \langle \Xi | c_{-i} c_0 c_1|0\rangle,
\;\;\;\;\;\;\;
\tilde{S}_{0i} = -\langle \Xi | c_{-i} c_{-1} c_1|0\rangle, 
\;\;\;\;\;\;\;
\tilde{S}_{1i} = \langle \Xi | c_{-i} c_{-1} c_0|0\rangle. 
\ee
To determine the correlation functions on the
left hand side of the above equations we require the knowledge of the
following similarity transformations.
\bea
U_{W_\infty} \alpha^\mu_{m}U_{W_\infty}^{-1} 
&=& \oint \frac{dz}{2\pi i} z^{m}
W_{\infty}\circ\partial X (z). \\ \nonumber
&=& \sum_{n=m}^{\infty}{\cal A}_{mn} \alpha^{\mu}_n,
\eea
where 
\bea
W_\infty &=&\lim_{n\rightarrow\infty}\tan(\frac{2}{n} \tan^{-1} z),
\\ \nonumber
&=& \lim_{n\rightarrow \infty} \frac{2}{n} \tan^{-1} z.
\eea
The correlation functions given in \eq{cor1}, \eq{cor2} and \eq{cor3}
are invariant under the scaling $z\rightarrow \frac{nz}{2}$. 
Therefore it is 
sufficient to work with $W_\infty (z)=\tan^{-1}z $ for the purposes of
evaluating the similarity transformations.
Thus  ${\cal A}_{mn}$ is given by the expansion
\be
(\tan z)^m  = \sum_{n=m}^{\infty} {\cal A}_{mn}z^n.
\ee
The similarity transformation for the ghosts are obtained from
\bea
U_{W_\infty }b_i U_{W_\infty}^{-1} 
&=& \oint \frac{dz}{2\pi i} z^{i+1} W_\infty\circ
b(z), \\ \nonumber
&=& \sum_{j=i}^{\infty}{\cal B}_{ij} b_j.
\eea
Here ${\cal B}_{ij}$ is defined in terms of the expansion
\be
(\tan z)^{i+1} \cos^2 z = \sum_{j=i}^{\infty}{\cal B}_{ij} z^{j+1}.
\ee
Similarly the the matrix ${\cal C}$ is represents the transformation
of the $c$ ghost by $U_{W_\infty}$.
\bea
U_{W_\infty} c_s U_{W_\infty}^{-1} 
&=& \int \frac{dz}{2\pi i} z^{s-2} W_\infty\circ
c(z) \\ \nonumber
&=& \sum_{t=s}^{\infty}{\cal C}_{st} c_t
\eea
Again ${\cal C}_{st}$ is defined in terms of the expansion
\be
\label{matc}
(\tan z)^{s-2} \cos^{-4} z = 
\sum_{s=t}^{\infty}{\cal C}_{st} z^{t-2}.
\ee
In appendix A we have listed down a few components of the matrices
${\cal A}, {\cal B}$ and ${\cal C}$. It is easy to see that 
from the definition of these matrices given above they
are twist invariant. For example ${\cal A}_{mn} =0$ if $m+n$
is an odd number.

Substituting the matrices 
corresponding to the similarity transformation in \eq{cor1} we obtain
\bea
S_{mn} &=& \frac{1}{mn}\sum_{m'=1}^{n} 
m'{\cal A}_{(-m)m'}{\cal A}_{(-n)(-m')} \left({\cal C}_{(-1)(-1)}{\cal
C}_{00}{\cal C}_{11} \right), \\ \nonumber
&=& \frac{1}{mn}\sum_{m'=1}^{n} 
m'{\cal A}_{(-m)m'}{\cal A}_{(-n)(-m')}. 
\eea
Here we have used the fact that the matrices ${\cal A}$ and ${\cal C}$
are upper triangular. In the second line we have used ${\cal
C}_{(-1)(-1)}= {\cal C}_{00} = {\cal C}_{11} =1$ and $\langle c_{-1}
c_0 c_1\rangle =1$.  Since the matrix ${\cal A}$ is twist invariant
the width matrix $S$ is also twist invariant. We write down a few
components of the width matrix $S$. These can be evaluated by the
coefficients of the ${\cal A}$ matrix listed in appendix
A\footnote{Note that one can obtain the coefficients of the width matrix
listed in \cite{rsz2} from these coefficients by replacing 
$\alpha_m^\mu$ by $i\sqrt{m} \alpha_m^\mu$. This is the normalization
of the matter oscillators used in \cite{rsz2}. }.
\bea
S_{11} = -\frac{1}{3},  \;\;\;\;\; S_{22} = -\frac{1}{30}, \;\;\;\;\;
S_{13} = \frac{4}{45},  \\ \nonumber
S_{33} = -\frac{83}{2835}, \;\;\;\;\; S_{24} = \frac{16}{945},
\;\;\;\;\; S_{44} = \frac{109}{11340}.
\eea
For the width matrix of the ghost sector $\tilde{S}$ we have 
\bea
\tilde{S}_{-1i} = 
{\cal C}_{-i-1}, \;\;&\;&\;\; \tilde{S}_{0i} = {\cal C}_{(-i)0},
\\ \nonumber
\tilde{S}_{1i} = {\cal C}_{(-i)1} - {\cal C}_{(-i)(-1)}{\cal
C}_{(-1)1},
\;\;&\;&\;\;
\tilde{S}_{si} = \sum_{k= -1}^s {\cal C}_{(-i)k} {\cal
B}_{(-s) (-k)}.
\eea
Here $i\geq 2$ and $s\geq 2$. 
Using the above equations and the coefficients of the ${\cal C}$ and
${\cal B}$ matrices listed in appendix A, 
we find the following coefficients
\bea
\tilde{S}_{(-1)3} = \frac{1}{3}, \;\;\; \tilde{S}_{(-1)5} =
-\frac{7}{45} \;\;\;
\tilde{S}_{02} = \frac{2}{3}, \;\;\; \tilde{S}_{04}= -\frac{2}{15}
\;\;\;\tilde{S}_{13} = -\frac{1}{3}, \\ \nonumber 
\tilde{S}_{15} = - \frac{7}{45}, 
\;\;\; \tilde{S}_{22} = -\frac{29}{45}, \;\;\; 
\tilde{S}_{24}= \frac{128}{945},  \;\;\;
\tilde{S}_{33} = \frac{61}{189}, \;\;\;
\tilde{S}_{35} = -\frac{2176}{14175}.
\eea

The sliver state can also be constructed using the explicit
representation of the Witten vertex in terms of  the Neumann functions
\cite{gj,sam, ohta} as proposed in \cite{kp}. 
These width coefficients obtained above are useful in
comparing with this explicit construction of the sliver state.
The matter sector width matrix $S$ has been compared to the explicit
construction of the matter sector of the sliver state in \cite{rsz2}.
The difficulty in comparing the ghost sector is that the
representation of the Witten vertex with the Neumann functions is on
the ghost number two vacuum. This makes
the evaluation of the star product difficult for a ghost number zero
state like the sliver state. It is of interest to point out that the
ghost sector of the squeezed state constructed in \cite{kp} 
satisfies the Feynman-Siegel gauge and it has ghost number one,
unlike the sliver state which does not satisfy the Feynman-Siegel
gauge condition and has ghost number zero.

\section{The generalized gluing and resmoothing theorem}

It is clear that the sliver state consists of infinite number of levels
above the $SL(2, R)$ vacuum. It would be difficult to compute the
Witten vertex of excitations on the sliver state directly. 
A direct computation of the Witten vertex of three ghost
number one states on the sliver using the Neumann function
representation of the Witten vertex leads to a determinant of a 
matrix of infinite dimension. It is therefore useful to develop a
convenient method to compute the Witten vertex and the quadratic term.
The generalized gluing and resmoothing theorem 
\cite{lpp,sensh} provides a means to
evaluate these vertices.
In this section we  will 
review the algebraic statement of the theorem. Then we
will apply it to re-derive the product law of the wedge states.

\subsection{Statement of the theorem}
We will follow \cite{sensh} in the algebraic statement of the theorem
and its application.
Consider a set of 
conformal maps $f_1, \ldots ,f_n, g_1, \ldots , g_m, f$
and $g$ with the property that they are well defined and one-to-one
inside a disc around the origin. Let $f$ and $g$ leave the origin
fixed. The conformal field theory of our interest is defined in the
upper half plane. Therefore the functions $f_i, g_i, f, g$ map the
real axis to the real axis.
The generalized gluing and resmoothing theorem states that
\bea
\label{ggrt}
&  \sum_r \langle f_1\circ\Phi_{r1}(0)\ldots f_n \circ\Phi_{r_n}(0)
f\circ\Phi_r(0)\rangle 
\langle g_1\circ\Phi_{s1}(0)\ldots g_m\circ\Phi_{s_m}(0)
g\circ\Phi_r^c(0)\rangle \nonumber \\ \nonumber 
&= e^{cK} \langle F\circ f_1\circ\Phi_{r1}(0)
\ldots F\circ f_n\circ\Phi_{r_n}(0) G\circ
I\circ g_1\circ\Phi_{s_1}(0)\ldots G\circ I\circ g_m\circ
\Phi_{s_m}(0) \rangle. \\  
\eea
Here $\{|\Phi_r\rangle\}$ is a complete set of basis and $\{\langle
\Phi_r^c|\}$ is basis dual to it with the property
\be
\langle\Phi_r^c|\Phi_s\rangle = \delta_{rs}
\ee
$F$ and $G$ are two conformal  maps with
the following properties.
\begin{itemize}
\item $F$ is a conformal map which is well defined and one-to-one
outside a disc around the origin. It leaves leaves the point at 
infinity fixed.
\item $G$ is a conformal map which is well defined and one-to-one
around the origin $(z=0)$. It leaves the origin fixed.
\item The maps $F\circ f(z)$ and $G\circ I\circ g\circ I (z)$ 
are well
defined over an annulus around the origin in the $z$ plane and in this
region
\be
F\circ f(z) = G\circ I\circ g\circ I(z).
\ee
\item Let the region in the unit disc $|z|\leq 1$ be mapped to the
region $D_1^c$ under the map $f(z)(\equiv u)$.
Call the complement of this region $D_1$. Similarly
let the region outside the unit disc $|z|\geq 1$ be mapped to the
region $D_2^c$ under the map $g(z)\circ I(z) (\equiv v)$. Call the
complement of this region $D_2$. The image of $D_1$ under the map
$F(u) (\equiv w)$ and the image of $D_2$ under the map $G\circ I(v)
(\equiv w)$ are complements of each other in the $w$ plane. The maps
$F$ and $G\circ I$ are well defined in regions $D_1$ and $D_2$
(See Figure 1 for the various domains and ranges involved.)
\end{itemize}

\FIGURE
{
\epsfig{file=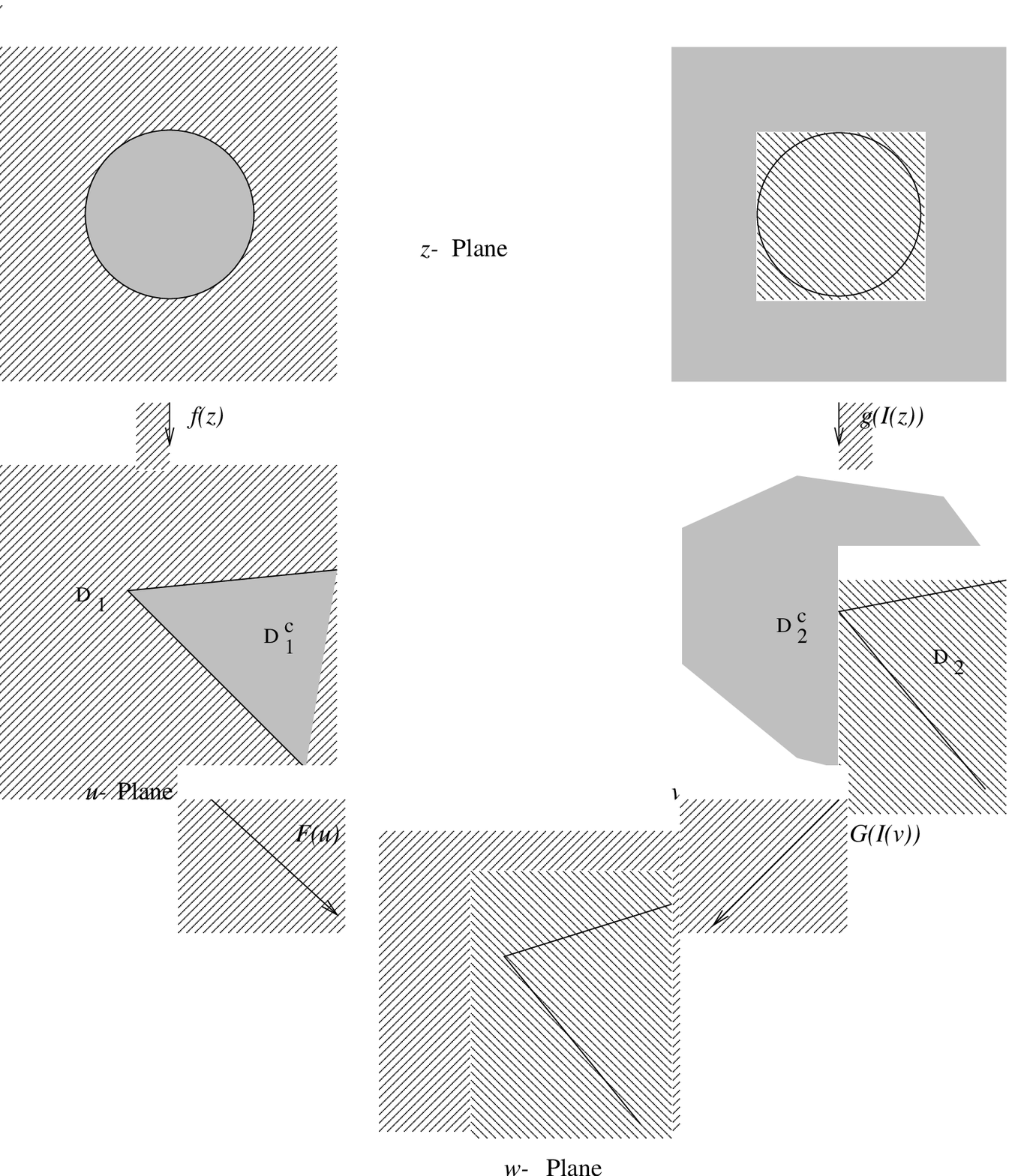}
\caption{Domains in the generalized gluing and resmoothing theorem}
}

\noindent
The factor $e^{cK}$ in \eq{ggrt} depends on 
the maps $f$ and $g$ and the central charge $c$ of the theory. For
the critical bosonic string theory in $d=26$ the central charge of the
ghost conformal field 
theory and the matter conformal field theory is zero.
Thus this factor becomes unity for the theory of our interest.

\subsection{Product law of wedge states}

In this section we will derive the product law of the wedge states
found in \cite{rz}. It was shown using using the geometrical
description of the generalized gluing 
and resmoothing theorem that $|n\rangle *|m\rangle =
|n+m-1\rangle$. We will re-derive this using the algebraic form of the
generalized gluing and resmoothing theorem. This will also help in
developing the method of evaluating the Witten vertex for excitations
above the sliver state.

We have seen that the wedge state is given by
\be
|n\rangle = U_{W_n}^\dagger|0\rangle = 
U_{I\circ W_n^{-1} \circ I} |0\rangle.
\ee
Inserting a complete set of basis 
$\sum_{r}|\Phi_r\rangle \langle \Phi_r^c|$ we get
\bea
\label{wstate}
|n\rangle &=& 
\sum_r|\Phi_r\rangle \langle 0|I\circ \Phi_r^c (0) U_{I\circ
W_n^{-1}\circ I } |0\rangle, \\ \nonumber
&=& \sum_r|\Phi_{r}\rangle \langle 0|U_{I\circ W_n\circ I} I\circ
\Phi_r^c (0) U_{I \circ W_n^{-1} \circ I} |0\rangle, \\ \nonumber
&=& \sum_r|\Phi_r\rangle
\langle 0 | W_n\circ \Phi_r^c(0) |0\rangle, 
\eea
Here we have used  $\langle 0| U_{I\circ W_n \circ I} =
\langle 0|$, this can be seen from \eq{sly}. In the third line we have
used the definition of conformal transformation on the  field
$\Phi_r^c(0)$ and the fact that correlation functions are invariant
under $SL(2, R)$ transformation. 
Using \eq{wstate} to express the wedge state $|n\rangle$ and
$|m\rangle$ in terms of a complete set of basis we can write the star
product of these states  with an arbitrary state $|\Phi\rangle$ as
\be
\label{law}
\langle V_3|\; | \Phi\rangle \otimes  |n\rangle \otimes | m \rangle =
\sum_{r,s}\langle f_1\circ\Phi (0) f_2\circ\Phi_r(0) f_3 \circ\Phi_s(0) 
\rangle \langle W_n\circ \Phi_r^c(0) \rangle 
\langle W_m\circ \Phi_s^c(0) \rangle 
\ee
Here $\langle V_3|$ stands for the Witten vertex.
The functions $f_1, f_2$ and $f_3$ are given by
\be
f_1 = M\left(e^{\frac{2\pi i}{3}} 
\left[\frac{1+iz}{1-iz}\right]^{\frac{2}{3}} \right), \;\;\;\;
f_2 = M\left(\left[\frac{1+iz}{1-iz}\right]^{\frac{2}{3}}
\right),  \;\;\;\;
f_3 = M\left(e^{\frac{4\pi i}{3}} 
\left[\frac{1+iz}{1-iz}\right]^{\frac{2}{3}} \right). 
\ee
where M stands for the $SL(2, C)$ transformation which maps the unit
circle to the real axis given in \eq{sl2r}. 

Now we use the generalized gluing and resmoothing theorem to sum over 
$r$ and $s$. 
As $f_2(0)=0$ and $W_n(0) =0$ we will first sum over $r$. 
From the statement of the theorem, we need to find the functions
$F_1$ and $G_1$ such that
\be
\label{combat}
F_1\circ f_2 (z) = G_1\circ I\circ W_n \circ I (z).
\ee
We look for these functions with the following ansatz
\be
\label{trans1}
F_1(u)  = M(e^{i\phi_1}(M^{-1}(u)^{\alpha_1})), \;\;\;\;\;\;
G_1\circ I (v)  = M((M^{-1}(v)^{\beta_1})).
\ee
The condition in \eq{combat} gives
\be
M\left[ 
e^{i\phi_1}
\left( \frac{1+ iz}{1-iz} \right)^{\frac{2\alpha_1}{3}} \right]
=
M\left[  
\left( \frac{1- i/z}{1+i/z} \right)^{\frac{2\beta_1}{n}} \right].
\ee
By comparing both sides we must have
\be
\label{match1}
\frac{2\alpha_1}{3} = \frac{2\beta_1}{n}.
\ee
The map $(1+iz/1-iz)^{2/3} = (\hat{u})^{\alpha_1}$ 
takes the region inside disc $|z|\leq
1$ to a wedge of angle $2\pi/3$. 
Let $D_1$ be the complement of this region.
$D_1$  has a wedge angle of $4\pi/3$ and is bounded by straight lines
passing thorough the origin at angles $\pi/3$ and $5\pi/3$.
Now
$e^{i\phi_1}\hat{u}^{\alpha_1}$ 
scales the wedge angle of the complement region to
$ 4\pi\alpha_1/3$. The phase rotates the region so that now it is
bounded by straight lines passing through $\pi\alpha_1/3 + \phi_1$ and
$5\pi\alpha_1/3 + \phi_1$
Call this region wedge $E_1$. 
Similarly the map $(1-i/z)/(1+i/z)^{2/n}=\hat{v} $ 
takes the outside of the unit disc
$|z|\geq 1$ to a wedge of angle $2\pi /n$. The complement of this
region is the wedge of angle $2\pi (n-1)/n$. Let this region be $D_2$
Under the scaling $\hat{v}^{\beta_1}=w$ the region $D_2$ is mapped to a
region $E_2$. This is a wedge of angle $2\pi\beta_1(n-1)/n$.
The condition for the maps $F_1$ and $G_1$ is that $E_1$ and $E_2$
should be the complements of each other in the $w$ plane. This gives
\be
\label{match2}
\frac{4\pi\alpha_1}{3} + 2\pi \frac{n-1}{n}\beta_1 = 2\pi.
\ee
Matching the phase angles so that the wedges 
$E_1$ and $E_2$ are complements of
each other we get
\be
\label{match3}
\frac{\pi\alpha_1}{3} + \phi_1 =   (\frac{\pi}{n} + \frac{2\pi
(n-1)}{n})\beta_1.
\ee
The solution of \eq{match1}, \eq{match2} and \eq{match3} are given by
\be
\alpha_1 = \frac{3}{n+1}, \;\;\;\;\;
\beta_1 = \frac{n}{n+1}, \;\;\;\;\;
\phi_1 = \frac{2\pi(n-1)}{n+1}.
\ee
Therefore by applying the generalized gluing and resmoothing theorem
for summing over $r$ in \eq{law} we obtain
\be
\label{law1}
\langle V_3| \; |\Phi\rangle \otimes |n\rangle \otimes | m \rangle =
\langle F_1\circ f_2 \circ \Phi(0) F_1\circ f_3 \circ \Phi_s(0) \rangle
\langle W_m \circ\Phi_s^c(0) \rangle.
\ee
Using the definition of $F_1$ we find
\bea
F_1\circ f_2 (z) &=& M \left[ e^{\frac{2\pi i n}{n+1} } 
\left( \frac{1+ iz}{1-iz} \right)^{\frac{2}{n+1}} \right] \\ \nonumber
F_1\circ f_3 (z) &=& M \left[ \left( \frac{1+
iz}{1-iz} \right)^{\frac{2}{n+1}} \right]
\eea
We use the same method to sum over $s$. We look for 
functions $F_2$ and $G_2$ such that
\be
F_2 \circ F_1 \circ f_3(z) = G_2\circ I \circ W_m \circ I (z).
\ee
We make the following antsaz for $F_2$ and $G_2$
\be
\label{trans2}
F_2(u) = M( e^{i\phi_2}(M^{-1}(u))^{\alpha_2} ), \;\;\;\;\;\;
G_2\circ I (v) =  M ( ( M^{-1} (v) ) ^{\beta_2} ). 
\ee
Using the similar methods of matching domains in the $w$-plane 
we obtain the equations for $\alpha_2, \beta_2, \phi_2$,
\bea
\frac{2\alpha_2}{n+1} &=& \frac{2\beta_2}{m}, \;\;\;\;\;
\frac{n}{n+1} 2\pi\alpha_2 + \frac{m-1}{m} 2\pi\beta_2 = 2\pi, \\
\nonumber
& &\frac{\pi}{n+1} \alpha_2 + \phi_2 = (\frac{\pi}{m} + 2\pi
\frac{m-1}{m} )\beta_2.
\eea
The solutions of the above set of equations are given by
\be
\alpha_2 = \frac{n+1}{n+m-1}, \;\;\;\;\;\;
\beta_2 = \frac{m}{n+m-1},    \;\;\;\;\;\;
\phi_2 = \frac{2\pi (m-1)}{n+m -1}.
\ee
The definition of $F_2$ allows us to find
\bea
F_2\circ F_1\circ f_1 (z) &=& M\left[  \left( 
\frac{1+iz}{1-iz} \right) ^{\frac{2}{n+m-1} } \right], \\ \nonumber
&=& W_{n+m-1} (z). 
\eea
We can now use the generalized gluing and resmoothing theorem 
to sum over $s$ in \eq{law1}. We obtain
\be
\langle V_3| |\Phi\rangle \otimes |n\rangle \otimes | m \rangle =
\langle W_{n+m-1} \circ \Phi (0) \rangle.
\ee
We have taken  
$\Phi$ to be  any string state.  $W_{n+m-1}$ is the conformal
transformation associated with the wedge state $|n+m-1\rangle$.
Thus we have proved
\be
|n\rangle * |m\rangle = |n+m -1\rangle.
\ee

\section{Excitations on wedge states}

In this section we define the excitations that we will consider
acting on the
sliver state. We will then use the generalized gluing and resmoothing
theorem to evaluate the Witten
vertex and the kinetic term for string field theory actions around the
tachyon vacuum.

\subsection{The similarity transformation}
As we have seen  in the section 2, the wedge states are defined by
\be
\langle n | = \langle 0| U_{W_n}, 
\;\;\;\;\;\;\;  |n\rangle =
U_{I\circ W_n^{-1}\circ I}|0\rangle.
\ee
$U_{W_n}$ has  definite representation in terms of
generators of the Virasoro group.
The wedge states are thus defined as a conformal transformation acting
on the $SL(2, R)$ vacuum.
To obtain creation and annihilation operators on the wedge state
$|n\rangle$ it is natural to define operators using 
the similarity transformation given below
\be
\label{simil1}
{\cal O}^{(n)} = U_{I\circ W_n^{-1}\circ I} {\cal O} U_{I\circ W_n
\circ I}, 
\ee
which implies
\be
{\cal O}^{(n)} |n\rangle = U_{I\circ W_n^{-1} \circ I} {\cal
O} |0\rangle.
\ee
Using this definition, for example
the  operator $c_{-m}^{(n)}$ for $m \geq-1$ act as creation
operators on the state $|n\rangle$. for $m <-1$ the operators
$c_{-m}^{(n)}$ annihilate the state $|n\rangle$. The
transformed operators give a convenient way of organizing the operators
as creation and annihilation operators over the wedge states. We will
denote ${\cal O}^{(\infty)}$ as $\tilde{{\cal O}}$. Thus
\be
\label{defsim}
\tilde{\cal O} = U_{I\circ W_\infty^{-1} \circ I} {\cal O} U_{I\circ
W_\infty \circ I}.
\ee
On the bra $\langle n|$ the similarity transformation for operators is
the BPZ conjugate of \eq{simil1}. It is given by
\be
({\cal O}^{(n)})^{\dagger} = U_{W_n}^{-1} {\cal O}^{\dagger} U_{W_n},
\ee
similarly
\be
\tilde{\cal O}^{\dagger} = U_{W_\infty}^{-1} {\cal O}^{\dagger}
U_{W_\infty}.
\ee
It is interesting to point out that the conservation laws of the wedge
state $\langle n|$ obtained in \cite{rz, rsz2} is due to the similarity
transformation of annihilation 
operators acting on the state $\langle n|$.
We demonstrate it for the sliver state. 
It is clear form the definition of $\tilde{c}_m$ that 
\be
\langle \Xi| \tilde{c}_{m}^{\dagger}  =0 
\;\;\;\;\;\hbox{  for  } m \leq -2.
\ee
Now let us express $\tilde{c}_{m}^{\dagger}$ 
in terms of the conventional
$c$'s. Using the definition of $\tilde{c}_{m}^{\dagger}$ we have
\bea
\tilde{c}_{m}^{\dagger} 
&=& U_{W_\infty}^{-1} c_{m} U_{W_\infty}, \\ \nonumber
&=& \oint \frac{dz}{2\pi i} z^{m-2} W_{\infty}^{-1}\circ c(z), \\
&=&\oint \frac{dz}{2\pi i} \frac{1}{(1+ z^2)^2}
(\tan^{-1} z)^{m-2} c(z).
\nonumber
\eea
It is sufficient to use $W_\infty^{-1}(z) = \tan(z)$. 
We have changed the
variable of integration in the third line.
Evaluating the integral for the cases $m =-2, -3, -4, -5, -6$
gives
\bea
\tilde{c}_{-2}^{\dagger} &=& c_{-2} -\frac{2}{3}c_0 + \frac{29}{45}c_2
-\frac{608}{945}c_4 + \cdots \\ \nonumber
\tilde{c}_{-3}^{\dagger} 
&=& c_{-3} -\frac{1}{3}c_{-1} + \frac{1}{3}c_1 -
\frac{61}{189} c_3 + \cdots \\ \nonumber
\tilde{c}_{-4}^{\dagger} 
&=& c_{-4} + \frac{2}{15} c_0 -\frac{128}{945} c_2 +
\frac{629}{4725}c_4 + \cdots \\ \nonumber
\tilde{c}_{-5}^{\dagger} 
&=& c_{-5} + \frac{7}{45} c_{-1} -\frac{7}{45} c_1 +
\frac{2176}{14175} c_3 + \cdots \\ \nonumber
\tilde{c}_{-2}^{\dagger} 
&=& c_{-6} -\frac{2}{35}c_0 + \frac{848}{14175} c_2
-\frac{1312}{22275} c_4 + \cdots
\eea
It is clear that these are the conservation laws obtained for the
sliver state in \cite{rsz2}.

\subsection{The Witten vertex for transformed operators}

In this section we derive a general formula for computing the Witten 
vertex of excited states defined in the previous section.
To be definite we evaluate
\be
S^{3}_{pnm}= \langle V_3| \, {\Omega}^{(p)} (0) |p\rangle\otimes
{\Psi}^{(n)}(0)|n\rangle \otimes 
{\Lambda}^{(m)}(0)|m\rangle. 
\ee
We write the excited state $|\Psi^{(n)}\rangle$ 
in terms of a complete set
of basis as follows.
\bea
\Psi^{(n)}|n\rangle &=& U_{I\circ W_n^{-1} \circ I} \Psi (0)
|0\rangle,
\\ \nonumber
&=& \sum_{r}
|\Phi_r\rangle \langle 0|I\circ \Phi_r^c(0) U_{I\circ W_n^{-1}
\circ I} \Psi (0) | 0\rangle, \\ \nonumber
&=& \sum_r|\Phi_r\rangle\langle 0|U_{I\circ W_n \circ I} 
I\circ \Phi_r^c(0)
U_{I\circ W_n^{-1}\circ I} \Psi(0) |0\rangle, \\ \nonumber
&=& \sum_r |\Phi_r\rangle \langle I\circ \Psi(0) W_n\circ \Phi_r^c(0) 
\rangle.
\eea
In the last step we have used the fact that correlation functions are
invariant under $SL(2, R)$ transformation $I(z)$. Using a similar
expansion in terms of a complete set of basis  for 
the states $\Omega^{(p)}|p\rangle$ and $\Lambda^{(m)}|m\rangle$ 
we obtain the following formula
for the Witten vertex $S^3_{pnm}$
\bea
S^{3}_{pnm} &=& 
\sum_{r, s, t} \left( 
\langle f_{1}\circ \Phi_t(0) f_2\circ \Phi_r(0) f_3 \circ \Phi_s(0)
\rangle \times \right.\\ \nonumber
&&\left. \langle I\circ \Omega(0)  W_{p} \circ \Phi_t^c(0) \rangle
\langle I\circ \Psi(0)  W_{n} \circ \Phi_r^c(0) \rangle
\langle I\circ \Lambda(0)  W_{m} \circ \Phi_s^c(0) \rangle \right).
\eea
Summing over $r$ and $s$ using the generalized gluing and resmoothing
theorem as stated in \eq{ggrt} twice we get
\be
\label{3vertexp}
S^{3}_{pnm} = \sum_t \langle F_2 \circ F_1 \circ f_1\circ \Phi_t(0)
F_2\circ G_1 \circ \Psi (0) G_2 \circ \Lambda (0)\rangle \langle
I\circ \Omega (0) W_{p} \circ \Phi_t^c (0) \rangle.
\ee
In the above equation the functions $F_1, G_1$ and 
$F_2, G_2$ are as found in \eq{trans1} and  \eq{trans2} respectively.
We now use the generalized gluing and resmoothing theorem again to sum
over t. For this we need to find functions $F_3$ and $G_3$ which
satisfy
\be
F_3 \circ F_2\circ F_1 \circ f_1(z) = F_3 \circ I\circ W_p\circ I(z).
\ee
We again make the antsaz
\be
F_3(u) = M( e^{i\phi_3} (M^{-1} (u) )^{\alpha_3}), \;\;\;\;\;
G_3\circ I(v) = M ((M^{-1} (v))^{\beta_3} ).
\ee
Using this antsaz and the methods described in the previous section 
we obtain the following equations for $\alpha_3, \beta_3, \phi_3$.
\bea
\frac{2\alpha_3}{n+m-1} &=& \frac{2\beta_3}{p}, \;\;\;\;\;
\left(\frac{n+m-2}{n+m-1}\right) 2\pi \alpha_3 + 
\left(\frac{p-1}{p} \right) 2\pi \beta_3 = 2\pi, \\ \nonumber
& &\frac{\pi \alpha_3}{n+m-1}  + \phi_3 =
\left(\frac{\pi}{p} + \frac{2\pi(p-1)}{p} \right) \beta_3.
\eea
The  solution of these set of equations is given by
\be
\alpha_3 = \frac{n+m-1}{n+m + p-3}, \;\;\;\;\;\;\;
\beta_3 = \frac{p}{n+m+p-3}, \;\;\;\;\;\;\;
\phi_3 = \frac{2\pi (p-1)}{n+m+p-3}.
\ee
Now  summing over $t$ in \eq{3vertexp} using the generalized gluing
and resmoothing theorem we get
\be
S^3_{pnm} = \langle G_3 \circ 
\Omega(0) F_3\circ F_2\circ G_1 \circ \Psi (0) F_3\circ
G_2\circ \Lambda (0)  \rangle,
\ee
where
\bea
\label{mwv}
F_3\circ F_2\circ G_1\circ (z) &=& M \left[ 
e^{\frac{i\pi n + 2\pi i(n+p-2)}{n+m +p -3}} 
\left( \frac{1+iz}{1-iz} \right) ^{\frac{n}{n+m +p-3} }
\right], \\ \nonumber
F_3\circ G_2 (z) &=& M \left[ 
e^{\frac{ i \pi m + 2\pi i (p-1) }{n+m +p -3} }
\left( \frac{1+iz}{1-iz} \right) ^{\frac{m}{n+m +p-3}} \right],
\\ \nonumber
G_3 (z) &=& M \left[ e^{\frac{i\pi p}{n+m +p -3}} \left(
\frac{1+iz}{1-iz} \right) ^{\frac{p}{n+m+p-3}} \right].
\eea
We have thus obtained the Witten vertex for excitations on wedge
states. Let us perform some simple checks on the vertex we have
obtained.
As the wedge state $|2\rangle$ is the vacuum state we expect
the vertex $S_{222}^3$ to be the ordinary Witten vertex. This can be
easily checked from \eq{mwv}. We see that \eq{mwv} reduces to the
ordinary Witten vertex for $n=m=p=2$.
The wedge state $|1\rangle$ is the identity of the string algebra.
This is not a normalizable state. This explains the reason that the
vertex $S^3_{111}$ is not well defined. We note that the vertex
$S^3_{nnn}$ for any $n$ has a cyclic symmetry in the fields $\Omega(0),
\Phi(0)$ and $\Lambda (0)$. This can be easily seen by the $SL(2, R)$
transformation $M\circ \exp( (2\pi i /3) \circ M^{-1}$ which permutes
these fields.
This symmetry is different from
the cyclic symmetry of the vertex in the 
fields $\Omega^{(p)}(0), \Phi^{(n)} (0),
\Lambda^{(m)} (0)$. The latter symmetry is due to the cyclic symmetry
of the Witten vertex. 

\subsection{The quadratic term for the transformed operators}

In this section we will calculate the quadratic term 
for the transformed
operators. Let the BRST operator at the tachyon vacuum 
be $Q$. The quadratic term in the string field
theory action is given by
\be
\label{oke}
S^2 = \langle I\circ \Phi(0) Q  \Phi(0) \rangle.
\ee
Let us now evaluate the kinetic term for states
$\Lambda^{(m)}(0)|m\rangle$ and $ \Psi^{(n)}(0)|n\rangle$. 
Substituting these states in the kinetic term we get
\bea
S^{2}_{mn} &=& \langle I\circ\Lambda(0) U_{W_m} Q U_{I\circ W_n^{-1}
\circ I}
\Phi(0) \rangle, \\ \nonumber
&=& \langle I\circ\Lambda(0) U_{W_m} 
U_{I\circ W_n^{-1}\circ I} \tilde{Q}
\Phi(0) \rangle,  \nonumber
\eea
where 
$\tilde{Q} = U_{I\circ W_n \circ I} Q U_{I\circ W_n^{-1}\circ I}$.
Inserting a complete set of states in the above correlation function we
get
\be
S^2_{nm} = \sum_r 
\langle I\circ\Lambda(0) W_n \circ \Phi_r(0) \rangle
\langle W_m\circ\Phi_r^c(0)  I \circ (\tilde{Q} \Phi(0)) \rangle
\ee
We will use the generalized gluing and resmoothing theorem to sum
over $r$. 
We need to find functions $F'$ and $G'$ such
that
\be
F'\circ W_n(z) = G'\circ I \circ W_m\circ I(z).
\ee
We make the usual anstaz
\be
F'(u) = M (e^{i\phi'} (M^{-1} (u))^{\alpha'} ), \;\;\;\;\;\;
G'(v) = M((M^{-1} (v))^{\beta'}).
\ee
Using the methods discussed in the previous section we obtain 
the following equations for $
\alpha', \beta'$ and $\phi'$
\bea
\frac{2\alpha'}{n} &=& \frac{2\beta'}{m}, \;\;\;\;\;
2\pi\alpha' (\frac{n-1}{n}) + 2\pi \beta' (\frac{m-1}{m}) = 2\pi,
\\ \nonumber
& &\frac{\pi \alpha'_2}{n} +\phi' = \left(2\pi\frac{m-1}{m} +
\frac{\pi}{m} \right)\beta'.
\eea
The solutions are given by
\be
\alpha' = \frac{n}{n+m-2},
\;\;\;\;\;\;\;
\beta' = \frac{m}{n+m-2}, 
\;\;\;\;\;\;\;
\phi'= 2\pi \frac{m-1}{n+m-2}.
\ee
So after summing over $s$ we obtain
\be
S^2_{nm} = \langle F'\circ I\circ\Lambda(0) G' \circ \tilde{Q}
G'\circ \Phi(0) \rangle
\ee
The functions in the vertex $S^2_{mn}$ is given by
\bea
\label{v2}
G'(z) &=&
M\left[ e^{\frac{i\pi m}{n+m-2} } \left(
\frac{1+iz}{1-iz} \right) ^{\frac{m}{n+m-2}} \right], \\ \nonumber
F'\circ I (z) &=& M 
\left[ e^{\frac{i\pi n+ 2\pi i(m-1)}{n+m-2} } \left(
\frac{1+iz}{1-iz} \right) ^{\frac{n}{n+m-2}} \right], \\ \nonumber
\eea
Note that the vertex $S^2_{mn}$ reduces to the ordinary kinetic term
given in \eq{oke} for $m=n=2$ as expected. The quadratic term
$S^2_{11}$ is not well defined 
due to the fact that the wedge state $|1\rangle$ is
not normalizable.

\section{Effective action for excitations on the sliver}

We have evaluated the Witten vertex and the quadratic term for
excitations on  wedge states. In this section we will 
derive the Witten vertex and the quadratic term for excitations on
the sliver state by taking an appropriate limit of the vertices
derived in section 4. We then discuss the discrete symmetries for the
quadratic term and the Witten vertex for excitations above the
sliver. The effective action is explicitly computed for level two
excitations on the sliver. 
Then we analyze the local maxima of the action.

\subsection{The sliver limit for vertices}

Consider the Witten vertex $S^3_{pnm}$, naively one would expect the
the vertex for excitations on the sliver is obtained by taking the
limit $p, n,m \rightarrow \infty$ in $S^3_{pnm}$. The resulting
expression is not well defined. This can be easily seen from \eq{mwv}. 
However the limit
$n\rightarrow\infty$ in $S^3_{nnn}$ is well defined. 
As discussed at the end of section 4.2 
this method of defining the limit also 
preserves the cyclic symmetry of the vertex for the fields $\Omega(0),
\Phi(0), \Lambda(0)$. This requirement of cyclic symmetry in these
fields is consistent with the expectation that the translational
invariant sliver state is unique.
We encounter a similar situation with the quadratic term. The naive
limit $m, n\rightarrow\infty$ in $S^2_{mn}$ is not 
well defined. But, the
limit $n\rightarrow\infty$ in $S_{nn}^2$ is well defined.  It
preserves the exchange symmetry of the quadratic term 
for fields $\Phi(0)$ and $\Lambda(0)$ which represent excitations 
on the unique sliver state.

\subsubsection {\bf The quadratic term }

As we have discussed above
the sliver limit for the quadratic term is obtain by
\bea
\tilde{S}^2 &=& \lim_{n\rightarrow \infty} S^{2}_{nn}, \\ \nonumber
&=& \langle g_1\circ \Phi(0) g_2\circ \tilde{Q}  g_2\circ \Lambda(0) 
\rangle.
\eea
After performing the $SL(2, R)$ transformation $M\circ \exp{(-i\pi /2)}
\circ M^{-1}$ on the vertex $\tilde{S}^2$ 
the functions $g_1(z), g_2(z)$ are given by
\bea
g_1(z) &=&  M\left[ e^{i\pi 
} \left( \frac{1+iz}{1-iz} \right)^{\frac{1}{2}} \right] 
 \\ \nonumber
&=& -\frac{2}{z} - \frac{1}{2}z + \frac{1}{8}z^3 - \frac{1}{8}z^5 +
\cdots, \\ \nonumber
g_2(z) &=& M\left[
\left(
\frac{1+iz}{1-iz} \right)^{\frac{1}{2}} \right] , \\ \nonumber
&=& \frac{1}{2}z -\frac{1}{8}z^3 + \frac{1}{16}z^5 - \frac{5}{128}z^7
\cdots.
\eea
We now define $\tilde{Q}$ in the sliver limit.
\be
\tilde{Q} = 
\lim_{n\rightarrow \infty}U_{I\circ W_n\circ I} \circ Q 
U_{I \circ W_n^{-1} \circ I}
\ee
In the sliver limit we can replace the operator $U_{W_n}$ by
$U_{W_\infty}$  where $W_\infty(z)= \tan^{-1}(z)$ \cite{rz}. 
The operator $Q$
is hermitian, therefore $I\circ Q= -Q$. Now we will discuss the
specific cases for $Q=Q_{BRST}, Q= c_0$ and $Q= c_0 + (c_2 +
c_{-2})/2$.

\vspace{1ex}
\noindent
{\bf Case 1. $Q= Q_{BRST}$}
\vspace{1ex}

\noindent
As the BRST current  is a primary with weight one it commutes with the
operator $U_{W_\infty}$. Therefore we get
\be
\tilde{Q}= Q_{BRST}.
\ee
Furthermore we also have the result $g_{2}\circ Q_{BRST} = Q_{BRST}$.

\vspace{1ex}
\noindent
{\bf Case 2. $Q= c_0$}
\vspace{1ex}
 
\noindent 
Using the definition of $c_0$ and of the
\bea
\tilde {Q} &=& -I\circ \int \frac{dz}{2\pi i} z^{-2} W_{\infty}\circ
c(z), \\ \nonumber
&=& \sum_{t=0}^{\infty} {\cal C}_{0t} c_{-t},
\eea
where the coefficients ${\cal C}_{0t}$ are defined in \eq{matc}. We
have use the fact that the ${\cal C}_{0t}$ is twist invariant and
$I\circ c_{2k} = - c_{-2k}$. 
We now need to evaluate $g_2\circ\tilde{Q}$. Writing the contour
integral representation for $c_t$'s we get
\be
g_2\circ \tilde{Q} = \sum_{t=0}^{\infty}{\cal C}_{0t}\oint
\frac{dz}{2\pi i}
\frac{g_2\circ c(z)}{z^{t+2}}. 
\ee
Here the contour is around the origin $z=0$. Now we perform  the
conformal transformation and change the variable of integration from
$z\rightarrow \tan(2\tan^{-1} z)$. We obtain the following expression
for $g_2\circ\tilde{Q}$
\bea
\label{csum}
g_2\circ\tilde{Q}
&=& \sum_{t=0}^\infty{\cal C }_{0t} \oint\frac{dz}{2\pi i}
\left[ 
\frac{1}{\tan(2\tan^{-1} z )}  \right]^{t+2}
\frac{4 c(z)}{(1+z^2)^2 \cos^4(2\tan^{-1} z)}, \\ \nonumber
&=& \sum_{t=0}^\infty{\cal C}_{Ot} \oint\frac{dz}{2\pi i} {\cal
V}_{c_0} (z),
\eea
where
\be
\label{cur}
{\cal V}_{c_0}(z) = \sum_{t=0}^\infty {\cal C}_{0t} 
\left[ \frac{1}{\tan(2\tan^{-1} z )}  \right]^{t+2}
\frac{4 c(z)}{(1+z^2)^2 \cos^4(2\tan^{-1} z)}.
\ee
We have to be careful about the contour in \eq{csum}. 
The contour splits into two contours. One is counter clockwise around
the origin and the other is counter clockwise around infinity.
This is because
both $\tan(2\tan^{-1} 0)$ and
$\tan(2\tan^{-1} \infty)$ is zero. We now specify the 
prescription for choosing the contour.
The correlation function $\tilde{S}^2$  for any two fields before
performing the contour integral is a function of $z$. This function is
defined below.
\bea
C(z)&=& \langle g_1\circ\Phi(0)c(z) g_2\circ\Lambda(0) \rangle,
\\ \nonumber
&=& \sum_n C_n z^n.
\eea
In the second line we have written down the Laurent series expansion of
$C(z)$. 
Substituting this expansion in \eq{csum} we get
\be
g_2\circ\tilde{Q} = \sum_{t=0}^\infty \oint \frac{dz}{2\pi i} 
\left[ 
\frac{1}{\tan(2\tan^{-1} z )}  \right]^{t+2}
\frac{4 }{(1+z^2)^2 \cos^4(2\tan^{-1} z)} \left( \sum_{n} 
C_n z^n \right)
\ee
It is easy to show that for the 
linear term and the
terms in the above series with
$t+1 \geq n$ and $t<-n$ and $n=1, 2, \cdots$,
the contour at infinity and the origin give the same result.
So we have to detail the prescription for $t+1<n$ and $t>-n $ with $
n= 1, 2, \cdots$.  
The prescription we adopt is for terms with positive powers of $z$ the
integral is chosen around the origin. For terms with negative powers of
$z$ the contour is at infinity. 
This definition of the contour defines a BRST
operator of the type $\sum_{n=0}^{\infty} a_{2n} {\cal Q}_{2n}$,
where ${\cal Q}_{2n} = (c_{2n} + c_{-2n})$
and $a_n$ are constants. This BRST operator is
Hermitian and and satisfies the required properties to
maintain gauge invariance of the cubic action \cite{rsz1}.
The expression for $g_2\circ \tilde{Q}$ in \eq{csum} consists of an
infinite sum. One can see from \eq{coefc} in appendix A the
coefficients ${\cal C}_{0t}$ decrease.
Evaluating the quadratic term $\tilde{S}^2$ for specific
fields of level $0$ and level $2$ 
shows that this series is a Liebnitz series which guarantees its
convergence. We demonstrate this convergence by evaluating the sum
till various values of $t_{max}$ for fields $\Phi(0) = \Lambda (0) =
c(0)$.
\vspace{1ex}
\be
\begin{array} {c  c c c c c c c c} 
\hline
t_{max} \;\;& \;0 \;& \;2\;& \;4\;& \;6\;& 
\;8\;&\; 10\;& \;12\;& \;14 \\
\hline
\\
\langle g_1\circ c(0) g_2(\circ \tilde{Q} c(0)  \rangle\;\;
& \;4 \;& 
\;6.666 \;& 
\;6.133 \;& 
\;6.302 \;& 
\;6.243 \;& \;6.264 \;& 
\;6.257 \;& \;6.260 \\
\\
\hline
\end{array}
\ee
\vspace{1ex}
It is easy to see that the series converges rapidly.

\vspace{1ex}
\noindent
{\bf Case 3. $Q= c_0 + (c_{2} +c_{-2})/2$}
\vspace{1ex}
\noindent
Using similar methods as in the case of $Q= c_0$, $g_2\circ \tilde{Q}$
is given by
\be
g_2\circ \tilde{Q} = \oint \frac{dz}{2\pi i} \left( {\cal V}_{c_0} (z)+
\frac{1}{2} ({\cal V}_{c_2} (z)+ {\cal V}_{c_{-2}} (z)) \right) .
\ee
The current ${\cal V}_{c_0}$ is given in \eq{cur}, while the 
currents ${\cal V}_{c_{2}}$ and ${\cal V}_{c_{-2}}$ are given by
\bea
\label{csum2}
{\cal V}_{c_2}(z) &=& \sum_{t=2}^{\infty}{\cal C}_{2t}
\left[ 
\frac{1}{\tan(2\tan^{-1} z )}  \right]^{t+2}
\frac{4 }{(1+z^2)^2 \cos^4(2\tan^{-1} z)}, \\ \nonumber
{\cal V}_{c_{-2}}(z) &=& \sum_{t=-2}^{\infty}{\cal C}_{(-2)t}
\left[ 
\frac{1}{\tan(2\tan^{-1} z )}  \right]^{t+2}
\frac{4 }{(1+z^2)^2 \cos^4(2\tan^{-1} z)}. 
\eea
The coefficients ${\cal C}_{-2t}$ are defined in \eq{matc}.
The prescription for the contour is the same as in the case for the
current ${\cal V}_{c_0}$. The series resulting from the above sum are
also Liebnitz series for fields till level $2$. They converge rapidly.
For an accuracy up to $3$ places in decimal it is sufficient to retain
$11$ terms in the series.

\subsubsection {\bf The Witten vertex}

\noindent
The sliver limit for the Witten vertex is obtained by
\bea
\tilde{S}^{3} &=&\lim_{n\rightarrow \infty} S^3_{nnn}, \\ \nonumber
&=& \langle 
\tilde{f}_1\circ \Omega(0)
\tilde{f}_2\circ \Psi(0)
\tilde{f}_3\circ \Lambda(0)
\rangle.
\eea
Taking the sliver limit for functions in \eq{mwv} we obtain
\bea
\tilde{f}_1 (z) &=& M \left[ e^{\frac{i\pi}{3}} \left( \frac{1+iz}{1-iz}
\right)^{\frac{1}{3}} \right], \\ \nonumber
&=& \frac{1}{\sqrt{3}} + \frac{4}{9}z + \frac{4}{27\sqrt{3}} z^2 -
\frac{28}{243} z^3 - \frac{20}{243\sqrt{3}} z^4 + \cdots, \\ \nonumber
\tilde{f}_2 (z) &=& M \left[ e^{\frac{i5\pi}{3}} 
\left( \frac{1+iz}{1-iz}
\right)^{\frac{1}{3}} \right], \\ \nonumber
&=& -\frac{1}{\sqrt{3}} + \frac{4}{9}z - \frac{4}{27\sqrt{3}} z^2 -
\frac{28}{243} z^3 + \frac{20}{243\sqrt{3}} z^4 + \cdots, \\ \nonumber
\tilde{f}_3 (z) &=& M \left[ e^{i\pi} \left( \frac{1+iz}{1-iz}
\right)^{\frac{1}{3}} \right], \\ \nonumber
&=& -\frac{3}{z} -\frac{8}{9}z + \frac{56}{243} z^3
-\frac{776}{6561}z^5 + \frac{4456}{59049} z^7 \cdots .
\eea
Note that it is easy to see that the vertex $\tilde{S}^3$ has cyclic
symmetry. The $SL(2, R)$ transformation $M\circ\exp({2\pi i/3})\circ
M^{-1}$ cyclically permutes the terms in the vertex.

\subsection{Discrete symmetries of the action} 

The vertices $\tilde{S}^2$ and $\tilde{S}^3$ inherits the same
discrete symmetries of the cubic string field theory. 
We show that the quadratic term vanishes for string states of
different world sheet parity when the BRST operator $Q$ is of even
parity.
The quadratic term  for excitations $\Phi(0)$ and $\Lambda(0)$ above
the sliver state is given by
\be
\label{qt}
\tilde{S}^2 = \langle g_1\circ \Phi(0) ( g_2\circ I\circ W_{\infty}
\circ I \circ Q) g_2 \circ \Lambda(0).
\ee
The maps $g_1, g_2, I $ have following property
\bea
M\circ\tilde{I} \circ M^{-1} \circ g_1(z) &=& 
g_1\circ P (z), \\
\nonumber
M\circ\tilde{I}  \circ M^{-1} \circ g_2(z) &=& 
g_2\circ P(z), \\ \nonumber
P\circ I (z) &=& I\circ P(z). 
\eea
Here $P$ and $\tilde{I}$ refer to the map $P(z) = -z$ and $\tilde{I} =
1/z$. $M\circ\tilde{I}\circ M^{-1}$ is a combination of world sheet
parity and $SL(2, R)$ transformation. 
Applying the
transformation $M\circ\tilde{I}\circ M^{-1}$ to each of the terms in
the correlation function $\tilde{S}^2$ leaves it invariant as the
$SL(2, R)$ vacuum is invariant. Using this transformation we can shift
the action of the parity to the fields $\Phi(0)$ and
$\Lambda(0)$ and to the current $W_\infty \circ Q$. Using the explicit
representation of $W_\infty \circ Q$ it is easy to see that it is left
invariant if $Q$ is of even parity. The net sign picked up is the
sum of parity of the two fields $\Phi(0)$ and $\Lambda(0)$.  Thus the
quadratic term vanishes for fields of different parity.

The quadratic term is also  symmetric under exchange of $\Phi(0)$ and
$\Lambda(0)$. This can be seen using the following properties of $g_1$
and $g_1$. 
\be
 I\circ   g_1(z) = g_2(z) 
\;\;\;\;\;\;  I \circ g_2(z) = g_1(z)
\ee
Furthermore from the explicit expressions for $g_2\circ \tilde{Q}$ 
in \eq{csum} and \eq{csum2} and the contour used in defining the
transformed BRST operator we obtain the result
$I\circ g_2\circ\tilde{Q} = -g_1\circ\tilde{Q}$. 
The correlation function $\tilde{S}_2(z)$ 
is invariant under the action $I$
on each of the fields. This exchanges the action of $g_1$ and $g_2$
with a  negative sign. Now interchanging the fields $\Phi(0)$ and
$\Lambda(0)$ picks up another negative sign. Thus the quadratic term
is symmetric in its two fields. There symmetry properties are borne
out by explicitly calculations till level 2.

The Witten vertex for excitations above the sliver
state has the following symmetry.
\bea
\tilde{f}_1(-z) &=& M\circ \tilde{I} \circ M^{-1} \circ
\tilde{f}_2(z),
\\ \nonumber
\tilde{f}_2(-z)&=& M\circ \tilde{I} \circ M^{-1} \circ \tilde{f}_1(z),
\\ \nonumber
\tilde{f}_3(-z) &=& M\circ \tilde{I} \circ M^{-1} \circ
\tilde{f}_3(z).
\eea
Using this symmetry one can show that the cubic vertex involving 
fields of ghost number one
with net even parity vanishes. This prohibits terms with one
parity even and two parity odd fields. It also prohibits terms with
three parity even fields. This twist symmetry of the cubic vertex 
for excitations above the sliver is the same as that of excitations
above the $SL(2, R)$ vacuum. 
As mentioned before the Witten vertex for excitations above the sliver
posses cyclic symmetry in its three fields. This can be seen by the
$SL(2, R)$ transformation $M\circ\exp(2\pi i/3)\circ M^{-1}$ which
cyclically permutes the terms in the vertex.

\subsection{The effective action}

In this section we will apply the Witten vertex and the quadratic
term derived for excitations on the sliver to evaluate the
effective action. We will find the gauge unfixed action for
fields representing excitations of level two on the sliver
state. The level of the excitation over the sliver state is defined by
the level of the operator ${\cal O}$ in \eq{defsim} creating the
excitation. The effective action is computed for two choices of the
BRST operator. $Q = c_0$ represents a class of BRST operators which 
do
not annihilate the identity of the string algebra and $Q= c_0 +
(c_2+c_{-2})/2$ represents the class which annihilates the identity.
The string field theory action at the tachyon vacuum is given by
\be
S(\Phi)= 
-\frac{K}{g_0^2} \left( \frac{1}{2} \langle \Phi |Q |\Phi\rangle +
\frac{1}{3} \langle \Phi, \Phi*\Phi \rangle \right).
\ee
Here $Q$ represents the BRST operator at the tachyon vacuum and $K$
stands for the unknown overall normalization constant of the BRST
operator. 
It must be possible to show from this action that one can obtain the
D25-brane solution. Using an antsaz that the solution factorizes into
the matter sector and the ghost sector the ratio
of various D-brane tensions 
to a high degree of accuracy in \cite{rsz2}. The translational
invariant solution in the matter sector used by \cite{rsz2} was the
matter sector of the sliver state. This motivates the use of the
complete sliver state including the ghost sector for the full
solution. As the string field has ghost number one we require a ghost
number one excitation on the sliver state. 
Assuming the factorization of the solution to the ghost sector
and the matter sector, the excitation can be chosen to be purely from
the ghost sector.
Thus the D25-brane will be a
represented by a local maximum in the effective action for these
excitations. Let the normalization of the D25-brane tension be given
by $1/(g^22\pi^2)$\cite{senu}. Then the potential is defined as 
\bea
V(\Phi) &=& -g^2 2\pi^2 S(\Phi), \\ \nonumber
&=& 2\pi^2 K\left( \frac{1}{2} \langle \Phi| Q |\Phi \rangle +
\frac{1}{3} \langle \Phi, \Phi*\Phi\rangle \right).
\eea
The definition of $V(\Phi)$ is chosen so that it equals $1$ for the
D25-brane solution. This will fix the unknown normalization constant
$K$. Because of the twist invariance of the action for excitations
above the sliver vacuum we can set all the odd level fields to zero.
We now analyse the effective action for
excitations above the sliver state for level $2$ fields.
The kinetic terms in  all of these effective potentials are evaluated
to $3$ decimal places of accuracy using \eq{csum} and \eq{csum2}.

\vspace{1ex}
\noindent
{\bf BRST operator $Q=c_0$} 
\vspace{1ex}

\noindent
At level $(0,0)$ the field is just the transformed tachyon
$t\tilde{c}_1$. The potential is 
\be
V^{(0,0)} =  2\pi^2K( 3.13 t^2 + \frac{81}{8\sqrt{3}} t^3).
\ee
The maximum of the potential at this level is attained at the
$ t= -0.357$. This gives $V^{(0,0)} = 2.624K$. 
Now we go over to level $(2, 4)$. 
The fields purely in the ghost sector are
\be
t\tilde{c}_1|\Xi\rangle + 
A\tilde{c}_{-1}|\Xi\rangle + B\tilde{b}_{-2} \tilde{c}_0 \tilde{c}_1
|\Xi\rangle.
\ee
We do not fix the Feynman-Siegel gauge as the sliver state itself
violates this gauge.
The effective potential to this level
is given by
\bea
V^{(2,4)} &=& 2\pi^2 K\left( 3.13t^2 + 1.956A^2 + 2.803B^2 + \right. \\
\nonumber
&&  4.694tA
 -5.933tB - 4.450 AB +   \\ \nonumber
& & \left. \frac{81}{8 \sqrt{3}} t^3 - \frac{69
\sqrt{3}}{8} t^2 A + \frac{175}{8 \sqrt{3}} tA^2 
 -  \frac{99\sqrt{3}}{8}t^2B + \frac{1085}{24 \sqrt{3}}
tB^2 - \frac{293}{ 4 \sqrt{3}} tAB  \right).
\eea
A local maximum is attained at $t= -0.496, A= .114,
B= .022$. The value of the tension at the potential at this point is
$V^{(2,4)} = 2.614K$. It is close to the level zero value.
Thus it seems that level truncation  
of the string field theory around the
tachyon vacuum using excitations on the sliver vacuum rather than the
$SL(2, R)$ vacuum seems to converge. It is interesting to point out
that level truncation of the string field theory with $Q=c_0$ using
excitations 
on the $SL(2, R)$ vacuum  in the Feynman-Siegel gauge pushed the
maximum to zero \cite{rsz1}.

\vspace{1ex}
\noindent
{\bf BRST operator $Q= c_0 + (c_2+ c_{-2})/2$}
\vspace{1ex}

We now repeat the calculation of the effective potential with the BRST
operator $Q=c_0 + (c_2+c_{-2})/2$. Only the kinetic term is modified.
At level $(0,0)$ we obtain the following potential
\be
V^{(0,0)} = 2\pi^2 K'\left( 4.254t^2 +
\frac{81}{3\sqrt{3}} t^3 \right).
\ee
Here $K'$ stands for the different normalization constant corresponding
to the BRST operator $Q= c_0 + (c_2 + c_{-2})/2$.
We obtain a local maximum at $t= -0.4852$. 
The potential at this point is $V^{(0,0)} = 6.589K'$. At level $2$ the
effective potential is 
\bea
V^{(2,4)} &=& 2\pi^2 K'\left( 4.254t^2 + 2.659A^2 + 3.767B^2 + \right.
\\ \nonumber
& & 6.381tA
- 8.021tB - 6.012AB + \\ \nonumber
& & \left. \frac{81}{8 \sqrt{3}} t^3 - \frac{69
\sqrt{3}}{8} t^2 A 
+  \frac{175}{8 \sqrt{3}} tA^2  
- \frac{99\sqrt{3}}{8}t^2B + \frac{1085}{24 \sqrt{3}}
tB^2 - \frac{293}{ 4 \sqrt{3}} tAB \right).
\eea
A local maximum occurs at $t= -0.675,  A= 0.158, B =
-0.0289$. The value of the potential at this point is $V^{(2, 4)} =
6.530K'$.  This also is close to the value of the potential obtained 
in 
the zeroth level approximation. Thus it looks like at least
to level two there is no  difference 
in the behaviour of the BRST
operators $Q=c_0$ and $Q=c_0+ (c_2 + c_{-2})/2$.

\section{Other excitations on the sliver state}

In this section we will discuss the difficulty with another 
choice of excitations on the sliver state.
Consider excitations of the form given by 
${\cal O}|\Xi\rangle$ where ${\cal O}$ are
operators built out of the ghost fields.
The conventional modes of the ghost fields 
cannot be organized as creation and annihilation operators over the
sliver state as $c_{m}|\Xi\rangle \neq 0$ for any value of $m$. It is
the same case with the $b$ ghost modes. We will show  the Witten
vertex for these excitations are not well defined. The Witten vertex
for these excitation can be derived using the generalized gluing
theorem and the methods discussed in section 4 and section 5. 
It is given by
\bea
\bar{S}^3 &=& \langle V_3 | \; \Omega|\Xi\rangle\otimes
\Phi|\Xi\rangle\otimes \Lambda|\Xi\rangle, \\ \nonumber
&=&\langle \bar{f}_1\circ \Omega \bar{f}_2 \circ \Phi
\bar{f}_3 \circ \Lambda \rangle.
\eea
where 
\bea
\bar{f}_1(z) &=& \lim_{n\rightarrow \infty} M\left[ 
e^{\frac{i\pi}{3}}\left( \frac{1+iz}{1-iz}\right)^{\frac{2}{3n-3}}
\right], \\ \nonumber
\bar{f}_2(z) &=& \lim_{n\rightarrow \infty} M\left[ 
e^{\frac{5i\pi}{3}}\left( \frac{1+iz}{1-iz}\right)^{\frac{2}{3n-3}}
\right], \\ \nonumber
\bar{f}_2(z) &=& \lim_{n\rightarrow \infty} M\left[ 
e^{i\pi}\left( \frac{1+iz}{1-iz}\right)^{\frac{2}{3n-3}}
\right]. \\ \nonumber
\eea
The presence of the finite phases in the definition of the
functions $\bar{f}_1(z), \bar{f}_2(z)$ and $\bar{f}_3(z)$ makes the
limit $n\rightarrow\infty$ ill defined. For example the Witten vertex
for the field $c_1|\Xi\rangle$ is given by
\bea
\bar{S}^3(c_1c_1c_1) &=& \langle 
\bar{f}_1\circ c_1 \bar{f}_2\circ c_1 \bar{f}_3\circ c_1
\rangle, \\ \nonumber
&=& \lim_{n\rightarrow\infty} \frac{81\sqrt{3}}{64} (n-1)^3.
\eea
The Witten vertex for these fields diverge as $n^3$. Thus these
excitations are not as well defined as the ones defined by 
the similarity transformation discussed earlier.

\section{Conclusions}

We have developed a method to analyse excitations on wedge states and
the sliver state. This is useful in exploring the solution of the
string field equations around the tachyon vacuum including the ghost
sector. 
We have used the gluing and resmoothing theorem to
re-derive the product law of wedge states explicitly. The theorem was
then used to construct the Witten vertex and the quadratic term for
excitations on wedge states and the sliver state. 
We verified that the
discrete symmetries for the effective action for excitations on the
sliver state was inherited form the cubic action.
We analyzed the gauge unfixed
effective action of ghost number one excitations on the
sliver till 
the second level for two choices of the BRST operator at the
tachyon vacuum. These excitations were purely in the ghost sector.
One expects the value of a 
local maximum of the effective potential to correspond to the tension of
the D25-brane upto an overall normalization constant
which depends on the choice of the BRST operator.
The choice $Q=c_0$ which represents a class of BRST
operators that 
annihilates the identity of the star algebra. For this choice  
we obtained that at level $(0,0)$ the value of the potential 
is $2.624K$ while at level $(2,4)$ it is $2.614K$. Here $K$ is the
unknown normalization constant corresponding to the BRST operator
$Q=c_0$.
The operator 
$Q= c_0 + (c_2+ c_{-2})/2$ represents a class of BRST operator
that annihilates the identity. The local maximum  at level $(0,0)$ is
given by $6.589K'$ and at level $(2,4)$ it is $6.530K'$.
$K'$ is the unknown normalization constant corresponding to the BRST
operator $Q= c_0 + (c_2 + c_{-2})/2$.
These results indicate that level truncation using 
purely ghost excitations on the sliver state seem to converge for both
choices of the BRST operator.
It also provides evidence for the  conjectured string field theory
actions around the tachyon vacuum.
It would be interesting to carry this analysis of the
effective action for
excitations purely  ghost sector 
over the sliver vacuum to a higher level.
We note that there are 
excitations on the sliver state for which the Witten vertex is not
well defined.

\acknowledgments

The author is especially grateful to Ashoke Sen for 
the many useful discussions
comments on the manuscript
and encouragement. He thanks Partha Mukhopadhyay and
Barton Zweibach for useful discussions. He is grateful to Spenta Wadia
for encouragement.
The work of the author is supported by NSF grant PHY97-2202.

\appendix
\section{Similarity transformation ${\cal A}, {\cal B}$ and ${\cal C}$}
The expression for the width matrix $S$ given in \eq{cor1} requires the
matrices ${\cal A}$ representing the similarity transformation of
$\alpha^\mu_m$ with the operator $U_{W_\infty}$. We list a few 
coefficients of the matrix ${\cal A}$ below.
\bea
U_{W\infty} \alpha_{-1}^{\mu} U_{W_\infty}^{-1} &=&
\alpha^\mu_{-1}
-\frac{1}{3}\alpha^\mu_{1}
-\frac{1}{45}\alpha^\mu_{3}
-\frac{2}{945}\alpha^\mu_{5}
-\frac{1}{4725}\alpha^\mu_{7}+ \cdots \\ \nonumber
U_{W\infty} \alpha_{-2}^{\mu} U_{W_\infty}^{-1} &=&
\alpha^\mu_{-2}
-\frac{2}{3}\alpha^\mu_{0}
+\frac{1}{15}\alpha^\mu_{2}
+\frac{2}{189}\alpha^\mu_{4}
+\frac{1}{675}\alpha^\mu_{6}
+\frac{2}{10395}\alpha^\mu_{8}+ \cdots \\ \nonumber
U_{W\infty} \alpha_{-3}^{\mu} U_{W_\infty}^{-1} &=&
\alpha^\mu_{-3}
-\alpha^\mu_{-1}
+\frac{4}{15}\alpha^\mu_{1}
+\frac{1}{945}\alpha^\mu_{3}
-\frac{11}{4725}\alpha^\mu_{5}
-\frac{29}{51975}\alpha^\mu_{7}+ \cdots \\ \nonumber
U_{W\infty} \alpha_{-4}^{\mu} U_{W_\infty}^{-1} &=&
\alpha^\mu_{-4}
-\frac{4}{3}\alpha^\mu_{-2}
+\frac{26}{25}\alpha^\mu_{0}
-\frac{64}{945}\alpha^\mu_{2}
-\frac{19}{2835}\alpha^\mu_{4}
-\frac{4}{22275}\alpha^\mu_{6} + \cdots 
\eea
The matrix ${\cal C}$ represents the similarity transformation of the
ghost modes $c_m$ by the operator $U_{W_\infty}$. 
This is required to evaluate the width matrix
$\tilde{S}$ of the ghost sector given in \eq{cor2} and \eq{cor3}.
We write down a few coefficients of ${\cal C}$ below.
\bea
\label{coefc}
U_{W_\infty} c_2 U_{W_\infty} ^{-1} &=&
c_2 + 2c_4 + \frac{7}{3} c_6 + \frac{94}{45} c_8 + \cdots  \\ \nonumber
U_{W_\infty}c_1 U_{W_\infty}^{-1}&=& 
c_1 + \frac{5}{3} c_3 + \frac{74}{45} c_5 +
\frac{239}{189} c_7 +\cdots\\ \nonumber
U_{W_\infty}c_0 U_{W_\infty}^{-1}&=& 
c_0 + \frac{4}{3} c_2 +\frac{16}{15} c_4
+\frac{128}{189} c_6 + \frac{256}{675} c_8 \cdots \\ \nonumber
U_{W_\infty}c_{-1}U_{W_\infty}^{-1} 
&=& c_{-1} + c_1 + \frac{3}{5} c_3 + \frac{274}{945}
c_5 + \frac{599}{4725} c_7 + \cdots \\ \nonumber
U_{W_\infty}c_{-2}U_{W_\infty}^{-1} 
&=& c_{-2} + \frac{2}{3} c_0 + \frac{11}{45}c_2 +
\frac{62}{945} c_4 + \frac{41}{2835} c_6 + \frac{62}{22275} c_8 +
\cdots \\ \nonumber
U_{W_\infty}c_{-3}U_{W_\infty}^{-1} 
&=& c_{-3} + \frac{1}{3} c_{-1} -\frac{31}{945} c_3
-\frac{41}{2835} c_5 - \frac{31}{7425} c_7 + \cdots \\ \nonumber
U_{W_\infty}c_{-4}U_{W_\infty}^{-1} 
&=& c_{-4} -\frac{2}{15} c_0 - \frac{8}{189} c_2
-\frac{1}{225}c_4 + \frac{8}{6237} c_6 + \frac{2218}{2764125} c_8 +
\cdots \\ \nonumber
U_{W_\infty}c_{-5}U_{W_\infty}^{-1}
&=& c_{-5} -\frac{1}{3} c_{-3} -\frac{7}{45} c_{-1} +
\frac{176}{14175} c_3 + \frac{53}{13365} c_5 +
\frac{17917}{30405375}c_7 + \cdots \\ \nonumber
\eea
We also need the similarity transformation of the modes of the $b$
ghost by $U_{W_\infty}$. These are 
listed below.
\bea
U_{W_\infty} b_{-2} U_{W_\infty}^{-1}
&=& b_{-2} -\frac{4}{3} b_0 + \frac{29}{45} b_2
-\frac{128}{945} b_4 + \cdots \\ \nonumber
U_{W_\infty} b_{-3} U_{W_\infty}^{-1}
&=& b_{-3} -\frac{5}{3} b_{-1} +\frac{16}{15} b_1
-\frac{61}{189}b_3 + \frac{31}{675} b_5 + \cdots \\ \nonumber
U_{W_\infty} b_{-4} U_{W_\infty}^{-1}
&=& b_{-4} -2 b_{-2} + \frac{8}{5} b_0 -\frac{608}{945}
b_2 + \frac{629}{4725} b_4 + \cdots
\eea

\section{Conformal transformations}
To evaluate correlation function with operators 
non-primary $c_0$, $c_{-1}$ and $b_{-2}c_{0} c_1$ we need the
following conformal transformations respectively.
\bea
f\circ\partial c(z) &=& -\frac{f^{\prime\prime}(z)}{(f^{\prime 2})(z)}
c(f(z)) + \partial c(f(z)), \\ \nonumber
f\circ\frac{\partial^2 c(z)}{2} &=&
\left[2\frac{(f^{\prime\prime}(z))^2}{(f^{\prime}(z))^3} 
 - \frac{f^{\prime\prime\prime}(z)}{(f^{\prime}(z))^2}\right] 
 \frac{c(f(z))}{2}
 -\frac{f^{\prime\prime}(z)}{f^\prime} \frac{\partial
 c(f(z))}{2}  \\ \nonumber
&+& f^\prime(z) \frac{\partial^2 c(f(z))}{2}, \\ \nonumber
 \\ \nonumber
f\circ b\partial c c(z) &=& f^{\prime}(z) b\partial c c(f(z))
+  \left[ \frac{2}{3} 
\frac{f^{\prime\prime\prime}(z)}{(f^{\prime}(z))^2}
-\frac{1}{4} \frac{(f^{\prime\prime}(z))^2}{(f^\prime(z))^3} \right]
c(f(z)) \\ \nonumber
&-&\frac{3}{2} \frac{f^{\prime\prime}(z)}{f^\prime(z)} \partial c(f(z)).
\eea

\end{document}